\documentclass[12pt]{book}
\usepackage[dvips]{graphicx,color}
\usepackage{makeidx,hori,zon}

\def\ket#1{\left| #1\right\rangle}

\makeauthorindex

\BookTitle{Inflating Horizon of Particle Astrophysics and Cosmology}
\CopyRight{\copyright 2005 by Universal Academy Press, Inc.\ and Yamada Science Foundation}
\begin{document}

\pagenumbering{arabic}

\chapter{The Black Hole Information Paradox}
\author{%
Martin B EINHORN\\
{\it Kavli Institute for Theoretical Physics,
University of California, Santa Barbara, CA 93106-4030\\
meinhorn@kitp.ucsb.edu}}
%
%

\AuthorContents{M.~B.~Einhorn} 

\AuthorIndex{Einhorn}{M.~B.} 

\section*{Abstract}

After a brief reminscence about work with K. Sato 25 years ago, a discussion is given of the black hole information paradox.  It is argued that, quite generally, it should be anticipated that the states behind a horizon should be correlated with states outside the horizon, and that this quantum mechanical entanglement is the key to understanding unitarity in this context.  This should be equally true of cosmologies with horizons, such as de~Sitter space, or of eternal black holes, or of black holes formed by gravitational collapse.

\section{Reminiscence on a collaboration.}

Several people have asked me how Katsu and I came to collaborate\cite{Einhorn:1980ik}.  As this conference is an opportunity to pay tribute to Katsu on his 60th birthday, perhaps I may be permitted some reflections on our truly serendipitous meeting.  First, a little background:  thinking back to 1979, magnetic monopoles were a hot topic because it had been realized that every grand unified theory (GUT,) such as SU(5), contained monopoles which would be essentially stable at scales below the unification scale.  The question arose about what their density should be today as relics of their production during the early universe\cite{Preskill:1979zi}.  For this, an estimate of their production during the early universe was required.  Stein, Toussaint, and I\cite{Einhorn:1980ym} had come to the conclusion that, if the universe underwent a second-order phase transition as the temperature cooled below the GUT scale, there would very likely be far too many relic monopoles around today.  Guth and Tye\cite{Guth:1979bh} had reached the same conclusion independently.  (As an aside, the two groups had submitted the papers to PRL with a request that they be published back-to-back.  Both were rejected!  How they came to be published separately is yet another interesting tale.  Some of this story is told in Guth's popular account\cite{Guth1}, but you may come away with the impression that, thereafter, I dropped off the edge of the earth.  Actually, I went to Copenhagen to spend a year visiting NORDITA and, in those pre-internet days, Alan and I lost touch for a while.

Shortly after arriving in January, 1980, I gave a seminar summarizing the results of this work and concluded by stating that I wanted next to understand was what happened if the phase transition were strongly first-order.   After the seminar concluded, Katsu came up and showed me a preprint that he had just written entitled ``First Order Phase Transition Of A Vacuum And Expansion Of The Universe."  (Although it was submitted in February, 1980, it was only published the following September.)  Katsu was half-way through spending a year in Copenhagen at the invitation of Chris Pethick, who had been impressed by Katsu's earlier work, discussed elsewhere in these proceedings.

You can imagine my surprise upon reading this; it contained much of what I had set out to learn about general relativity in this context.   This was wonderful, because he came at this from an astrophysics/cosmology  background, whereas I came from a particle physics background and was still learning cosmology.  Sato's analysis had been made generically, motivated by a first-order phase transition in the Weinberg-Salam model.  He wasn't so familiar as I with Grand Unified Theories and not at all with the monopole puzzle, but he didn't need much convincing to become persuaded that this was the more natural context for applying this cosmology.

This was easier said than done, and it wasn't clear how that was to be married to the scenario of a GUT broken as the universe cooled and, in particular, what effects this might have on the monopole density and how the universe might appear after the phase transition.  Guth and Tye were also led in the same direction, and, I later learned, had also realized that an exponential expansion of the universe would occur that might dilute the monopole density significantly greater than in the second-order case.  

By mid-March, when I traveled to Erice to lecture on these things at the Europhysics Conference on Unification of the Fundamental Interactions\cite{Einhorn:1980rt},
 we had developed a fairly clear picture that the exponential expansion would dilute the monopole density sufficiently, but it was much more difficult to see whether the universe would be left sufficiently homogeneous.  I recall when I gave my talk, Lenny Susskind, who had just come from Stanford, volunteered that I had just described  ``Guth's Cosmology."  No surprise, Alan was at SLAC and had been led down the same path.  The pictures that we had independently arrived at came to be called ``old inflation," and we all concluded that it left the universe too grainy to work.  To his enduring credit, Alan realized that the exponential expansion might solve some other cosmological issues, and, as they say, the rest is history.   Nevertheless, the inflationary universe emerged from the monopole problem, not from those other issues.  

Unfortunately, Katsu had to return to Japan at the end of June, and in those days, collaboration at a distance was much harder than now, so we weren't able to continue working together.  Sato and I submitted our work in July.  Others, more stubborn than we, persisted in developing alternate models of inflation.  I was then as now (and like many others) terribly bothered by the fine-tuning of the cosmological constant inherent in all such models and felt that, to make progress, we needed to understand naturalness.  Remarkably, it seems that Nature doesn't care!  

It is a great pleasure to be here to celebrate Katsu's 60th birthday and the 25th anniversary of our collaboration.  I have always regarded our meeting as a most fortuitous happenstance.  It is my pleasure to wish you ``otanjoubi omedetou gozaimasu."

\section{Introduction to the Black Hole Information Paradox}

I now turn to the black hole (BH) information paradox.
In 1975, Hawking showed that, because of quantum effects, it appears to a stationary observer that radiation is emitted from a region near the horizon of a black hole.\cite{Hawking:1974sw}  With the approximations he used, it seemed that the radiation was purely thermal.  Shortly thereafter, he realized that, if a BH eventually evaporated due to this mechanism of energy loss, it would challenge the basic tenets of quantum mechanics.  In the case of gravitational collapse, one may suppose that matter starts in a pure state, collapses to form a BH, which eventually evaporates leaving the universe in a thermal  mixed state.  However, unitary evolution in quantum mechanics implies that pure states evolve to pure states, so that that  somewhere in this process, there must be a breakdown of unitarity.\cite{HawkingBHI}.  This observation led to many attempts to circumnavigate this problem or to improve upon his approximations.  (For a brief review with many citations to the literature, see ref.~\cite{Page:2004xp}.)  This paradox is closely connected with Beckenstein's conjecture\cite{Bekenstein:1973ur} that the entropy of a BH is proportional to the area of its horizon rather than its volume, as might naively be expected in quantum field theory (QFT.)  This, together with corresponding laws of BH thermodynamics,  have  generally become accepted, at least at the semiclassical level.  Eventually, it suggests that, at a fundamental level when gravity is involved, QFT is a redundant description and that dynamics may be describable in terms of many fewer degrees of freedom than had been previously thought, a property that has been called the holographic principle\cite{'tHooft:1993gx, Susskind:1994vu}.

Motivated by Maldacena's treatment of a BH in AdS\cite{Maldacena:2001kr}, Hawking recently changed his mind and is now of the opinion that information is not lost via BH evaporation\cite{HawkingDublin, Hawking:2005kf}.   The AdS/CFT correspondence does carry with it the strong suggestion that the evolution of a BH is unitary.   The key element in Maldacena's view, which Hawking had overlooked, is the role of different topological configurations, but Barbon and Rabinovici\cite{Barbon:2003aq} showed that the situation is necessarily more complicated than Maldacena suggested.  Moreover, Hawking's  argument that, at the end of the day, gravitational collapse to a BH does not contribute to the S-matrix at infinity seems to throw the baby out with the bathwater, so his  widely publicized  ``concession" remains quite controversial.  Simply regarding a BH like another resonance in scattering amplitudes does seem to beg the question.  Thus, there remains much else to be understood.  Precisely, how can a black hole appear at one and the same time to be black body and yet be part of a pure state?  And how is that to be reconciled with the successful accounting in string theory of the BH microstates\cite{Strominger:1996sh,Maldacena:1997de,Gubser:1996de}?  

The information paradox is sharpest in the case of gravitational collapse from a pure state having no horizon to a BH followed by its eventual evaporation leaving an apparently thermal state.  The question is whether this process is like the burning of a book, in which there is no doubt that the information is encoded in the radiation even though it would be impossibly difficult to recover.  Thus, the issue is a matter of principle rather than of observation.  Unfortunately, the dynamics of the collapse followed by radiation are not well understood in a detailed way.   What has been better understood is the quantum character of stationary BHs, for which there is no immediate paradox.  To arrange for such a situation, one must artificially place sources at infinity so that the incoming energy precisely balances the radiation, leaving a stationary horizon.  This allows one to consider a quantum field in a fixed background.  We shall review this situation first and subsequently reflect on the case of gravitational collapse.

\section{Entanglement Entropy}

Our proposal is that BH entropy can always be identified with entanglement entropy.  (For the case of eternal BH's, this point of view is elaborated in ref.~\cite{BEY}.)  I would like to review entanglement entropy in detail, but, given space limitations, I must refer you to ref.~\cite{BEY} or to Feynman's lectures\cite{Feynmanstatmech}, whose notation I will closely follow.  The main point is as follows:  Suppose a system, which might be the entire universe, is in a pure state $\ket{\psi}.$  An observer who is confronted with a horizon cannot make measurements on the system beyond that.  Therefore, it is natural to express the state in a basis of states describing states within her causal sector $\ket{\phi_i}$ and states outside 
$\ket{\theta_r}$. These span Hilbert spaces that we will call ${\cal{H}}_1$ and ${\cal{H}}_2,$ respectively, and the full space of states is the Hilbert space ${\cal{H}}= {\cal{H}}_1\otimes {\cal{H}}_2.$   The general state is then written as 
\begin{equation}\label{purestate}
|\psi\rangle =\sum_{i,} C_{ir} |\phi_i \rangle_1 |\theta_r\rangle_2,
\end{equation}
 or, as a density matrix,   
$\rho=|\psi\rangle \langle\psi|.$  (Through this discussion, the indices may actually be vector indices, representing all the quantum numbers necessary to uniquely specify the basis states.)
Since $\rho$ corresponds to a pure state, $\rho^2=\rho.$ 
An observable for the observer outside the horizon corresponds to a Hermitian operator $A$ that acts within ${\cal{H}}_1,$ that is, 
whose domain and range are both in the subspace ${\cal{H}}_1.$  Then the expectation value of $A$ is
\begin{eqnarray}
\langle\psi|A|\psi\rangle&=&\sum C_{js}^*C_{ir}
\langle\theta_s|\langle\phi_j|A|\phi_i\rangle|\theta_r\rangle\\
&=&\sum(C C^\dagger)_{ij}\langle\phi_j|A|\phi_i\rangle\\
&=&{\rm Tr_1}\left[A\rho_1\right],\ {\rm where\ \ }\rho_1\equiv {\rm Tr_2}[\rho]=C C^\dagger.
\end{eqnarray}
In physical terms, an observer constrained to subspace ${\cal{H}}_1$ appears to be in a mixed state described by the density matrix $\rho_1.$  Diagonalizing $\rho_1,$ it can be expressed in a new basis $|i\rangle_1$  for ${\cal{H}}_1$  as 
\begin{equation}
\rho_1=\sum w_i|i\rangle_1{}_1\!\langle i|,\ {\rm with~eigenvalues\ } 0\le w_i\le1.
\end{equation}
The mixed state arises because of the correlations between ${\cal{H}}_1$ and ${\cal{H}}_2$ implied by the fact that, globally, the universe is in a pure state.  Even if the universe were in its vacuum state; the observer would appear to observe a distribution of excited states.  This is one of many such concepts that takes getting used to.

The entropy associated with the density matrix $\rho_1,$ called entanglement entropy, is defined as usual as
\begin{equation}
S_1=-\sum w_i \ln w_i = - {\rm Tr_1}[\rho_1\ln(\rho_1)].
\end{equation}
Similarly, an observer confined to ${\cal{H}}_2$ sees a density matrix 
\begin{equation}
\rho_2\equiv {\rm Tr_2}[\rho]= C^\dagger C.
\end{equation}
Now it is easy to show that the nonzero eigenvalues of $\rho_1$ and $\rho_2$ are the {\bf same} (even if they have different dimensions.)   In particular, the entanglement entropies of the two subspaces must be equal:
\begin{equation}
S_1=S_2.
\end{equation}
This simple observation has far-reaching consequences.

Finally, if there is equal likelihood to be in each state $|i\rangle_1,$ then $w_i=1/N_1,$ where $N_1$ is the dimension of ${\cal{H}}_1$.  Then one sees that $S_1=\ln N_1.$  In that case, the entanglement entropy coincides with the log of the number of microstates, which is the common definition of entropy for a thermal ensemble.

\section{Quantum Field Theory in Curved Backgrounds}

I cannot review here the nature of QFT in curved spacetime\cite{Birrell, Brout:1995rd}.  Let me recall a few salient facts:  
The Hilbert space of states is in general coordinate dependent, and there are alternative, frame-dependent definitions of the no-particle state.  Often there is no conserved Hamiltonian, so there is no analogue of the lowest-energy state or ground state naturally associated with the vacuum for QFT in Minkowski space.  Gravitational collapse is hard to discuss, but the entropy of eternal black holes or stationary spacetimes is somewhat easier.  Classic discussions compare QFT in an inertial frame in flat space (Minkowski) versus a constantly accelerating frame (Rindler), or the nature of a Schwarzschild BH in Schwarzschild coordinates as compared with Kruskal-Szekeres coordinates.  

Similarly, there are cosmologies such as de~Sitter spacetime, in which observers would observe radiation from the horizon, as well as more complicated cases, such as the Schwarzschild BH in asymptotically AdS space\cite{Maldacena:2001kr}.

All these situations have in common that there exists a stationary coordinate system with timelike Killing vector  in which the spacetime admits a bifurcating Killing horizon.  (See figure and references in \cite{BEY}.)   The Hilbert space may be built up as a product $\cal{H}_{\rm L}\times \cal{H}_{\rm R}$ of Hilbert spaces representing states associated with the ``Left" or ``Right" portion of a fixed time slice, with Hamiltonians $\rm{H_L}$ or $\rm{H_R}$ for each.  The sense of the time is reversed across the horizon, so the full Hamiltonian is $\rm{H=H_R-H_L}.$   A natural choice of the global vacuum state in such cases, consistent with the stationary metric, is the Hartle-Hawking state.  Typically, in these cases, the vacuum takes the form\cite{Israel:1976ur}
 \begin{equation}
\ket{0}=\frac{1}{\sqrt{Z}} \sum_i e^{-\frac{\beta E_i}{2}} \ket{E_i}_R\ket{E_i}_L,
\end{equation}
where $Z$ is the partition function,  $E_i$ is the eigenvalue of $H_R,$  and $\beta$ is the inverse temperature associated with the Hawking radiation.  For example, for the Schwarzschild BH, $\beta=8\pi G_NM_{BH},$ where $G_N$ is Newton's constant and $M_{BH}$ is the mass of the black hole.  This gives a thermal density matrix for $\rho_{\rm R}$ and an entropy equal to the Beckenstein-Hawking value, $S_{BH}=A/4G_N,$ where $A$ is the area of the horizon. 

Unfortunately, the case of eternal BH's, while illustrating the role of entanglement in BH entropy, does not shed much light on the nature of the information paradox.  It may help to have a more microscopic picture of the mechanism of radiation and the associated back reaction on the background metric.  While Hawking radiation is sometimes referred to as a tunneling process, the only quantitative development of this idea that I've seen is one by Parikh and Wilczek\cite{Parikh:1999mf}.  A BH that radiates is analogous to the decay of an unstable particle, $M\rightarrow M'+\gamma,$ 
in which a particle of mass $M$ decays into another particle of mass $M'$ plus a photon. These authors show that the transition can be calculated semiclassically in a particularly nice, stationary coordinate system that is nonsingular at the horizon (while remaining asymptotically flat.)  Energy is conserved however, so that $M=M'+\omega_\gamma,$   Thus, the black hole horizon correspondingly shrinks, in the Schwarzschild case, from $R=2G_NM$ to $R'=2G_NM'=2G_N(M-\omega_\gamma).$  They calculate the transition amplitude in WKB approximation.  Interestingly, they do not obtain a perfectly thermal spectrum except for sufficiently small $\omega_\gamma.$   Their microscopic description suggests that the continual decrease of entropy during radiation is perfectly consistent with unitarity when the backreaction on the background is taken into account.  

\section{Gravitational Collapse}

There have been precious few calculations of gravitational collapse that bear on the issue of the information paradox.  However, if one believes in quantum mechanics and unitarity, then it is easy to argue that the entropy of a BH so formed will be entanglement entropy.  Imagine starting in the distant past with a pure state of matter in a background having no horizon and only infalling matter, with none outgoing  $\ket{\psi, g_{\mu\nu}}.$  The matter, together with a self-consistently determined metric, evolves unitarily in time $U(t)\ket{\psi, g_{\mu\nu}}.$   At some point, the background develops a horizon with respect to a stationary observer outside.  Eventually, all of the original matter will have fallen inside the horizon  although there will remain outgoing radiation outside. Classically, the two regions are causally disconnected.  Since measurements are classical, no observer subsequently can determine all the properties of the system, so it is natural to describe the Hilbert space as a product of states inside and outside the horizon.  Globally, since the system is in a pure state, it will take the form of eq.~(\ref{purestate}).    This must remain true as a function of time, even as the collapse continues while radiation is emitted from the neighborhood of the horizon.  An exterior measurement during this period would necessarily involve a density matrix associated with the causal region of a given observer.  But the concept of a measurement  ``during this time" is already problematic, since one can only determine energy with limited accuracy in a finite time.  (Basically, the approximation is that the metric changes slowly compared with the duration of the measurement to some accuracy.)  We think eventually the BH so formed would evaporate.  Thus, the BH is like a resonance in particle physics, in which an initial state of matter forms a long-lived configuration that eventually decays back to ordinary matter.  Asymptotically, one only measures stable particles, ``from dust we were made, and to dust we shall return."\cite{Genesis3:19}

Perhaps one could describe the evolution of the metric quasi-statically and adiabatically.  At first, that sounds contradictory, since that suggests no entropy production.  However, globally, since the system remains in a pure state, there is nothing in principle to obstruct such an argument.

Recall that the entanglement entropy of each subsystem is always equal, so the question of the number of states inside and outside the BH seems not so relevant.  From this point of view, it is puzzling that  string theory calculations, counting black hole microstates, work\cite{Strominger:1996sh,Maldacena:1997de,Gubser:1996de}.
   Based on the observation in Section~1 (below eq.~(8)), the suggestion would be that, by the time string theory is relevant, we are dealing with strongly interacting gravity, in which all states in the available phase space are equally likely to be populated.   Some sort of ergodic theorem in which the time-averaged properties of such a system resemble ensemble averages would then be needed to reconcile these two points of view.  This is how statistical mechanics normally reconciles the behavior of a unique system with that of a collection of similar systems.  The brilliance of the argument in ref.~\cite{Strominger:1996sh} was the realization that, owing to supersymmetric nonrenormalization theorems, the counting could be done in the weak coupling regime.

\section{Summary and Conclusions}

In the light of our previous discussion, it is worth reflecting on various other attempts to reconcile the information paradox with the usual laws of physics.  One focus has been on the BH final state, whether there is a remnant left containing all the information that has fallen into the BH, or whether it completely evaporates.  This line of development is misdirected, because Hawking radiation appears to come from the horizon.  As the mass of the BH decreases, its horizon shrinks and its entropy diminishes.  So one must understand whether or not this process involves information loss.  The issue is not what happens at the very end.  

Similarly, a great deal of attention has been devoted to the question of the singularity inside.  This IS a very interesting question, and if it were resolved by a wormhole tunneling to a different classical spacetime, then certainly, not only information but also energy might disappear from our universe.  The AdS/CFT correspondence provides hope that this is not the outcome and suggests that, in the end, all will be well.  Regardless, this does not address the decrease of entropy in the initial stages discussed above.  It may however effect things near the very end.

What happens in the final stage of evaporation is anybody's guess, so here is mine.  I suspect that it is quite uneventful, since as the BH radiates, the mass decreases along with the entropy.  Eventually, it becomes very small with very few states left to be entangled.  (Recall $S_1=S_2$ always.)   The final stage ends with a whimper, not a bang.  

If the viewpoint elaborated here is correct, the BH information paradox will eventually be seen as similar to the EPR paradox\cite{Einstein:1935rr}.  On the one hand, the situation seems to violate the laws of physics. On the other hand, it clearly does not, but, because it is counterintuitive, it still is fascinating.  Thus, rather than a physics problem, it becomes a psychological problem, a bit like an optical illusion.

\section{Acknowledgement}

I would like to thank my coauthors of ref.~\cite{BEY} for extensive discussions about the  topics surveyed herein.   I also wish to thank the conference organizers for their hospitality and for the opportunity to return to Japan on this happy occasion.





\end{document}